# Transport properties of annealed CdSe nanocrystal solids

M. Drndic[1], M. Vitasovic[2], N.Y. Morgan[1], M.A. Kastner[1], M.G. Bawendi[2]

*Department of Physics[1] and Department of Chemistry[2],*

*Center for Material Science and Engineering,*

*Massachusetts Institute of Technology, Cambridge, MA 02139*

## Abstract

Transport properties of artificial solids composed of colloidal CdSe nanocrystals (NCs) are studied from 6 K to 250 K, before and after annealing. Annealing results in greatly enhanced dark and photocurrent in NC solids, while transmission electron microscopy (TEM) micrographs show that the inter-dot separation decreases. The increased current can be attributed to the enhancement of inter-dot tunneling caused by the decreased separation between NCs and by chemical changes in their organic cap. In addition, the absorption spectra of annealed solids are slightly red-shifted and broadened. These red-shifts may result from the change of the dielectric environment around the NCs. Our measurements also indicate that Coulomb interactions between charges on neighboring NCs play an important role in the tunneling current.

## I. INTRODUCTION

Electrons confined to nanometer-sized regions in semiconductors or metals have been the subject of intense study for many years. The quantization of charge and energy resulting from the confinement makes such regions, or quantum dots, behave like artificial atoms [1]. With recent advances in techniques for synthesizing large numbers of metallic and semiconductor nanocrystals (NCs), it is now possible to make large arrays of artificial atoms in which particle size, inter-particle separation and chemical composition are controlled [2]. This opens the possibility of creating artificial solids with tunable electrical and optical properties. For example, it has been demonstrated that a Langmuir film of silver NCs undergoes a metal-insulator transition as a function of the separation between NCs in the array [3].

Solids composed of metallic NCs have transport properties that are well described by a model of collective charge transport in disordered arrays [4]. Whereas the large number of conduction electrons in the metallic NCs provide substantial screening of long-range Coulomb interactions between electrons on different NCs, semiconductor NCs make possible the study of artificial solids in which such long-range interactions are very large. In these systems, transport measurements may show collective effects of charge ordering recently proposed by Levitov et al. [5].

So far, transport measurements on arrays of colloidal CdSe NCs separated by organic capping molecules have shown that these artificial solids are surprisingly insulating [6-8]. Long-range Coulomb interactions between charges on different NCs slow the dynamics of the injected electrons, an effect best described by an electron (Coulomb) glass model [9]. The current decays over extremely long time scales in response to a voltage step, and there is no steady-state current [6,7].

In order to increase the electrical current through the CdSe NC solids, the tunneling of electrons between NCs must be enhanced. This can be achieved by decreasing the physical separation and the height of the tunnel barriers between NCs in the array. The inter-dot separation is originally set by the length of the organic capping molecule [2]. One way to decrease the inter-dot separation, while keeping the NC size the same, is to anneal the NC solid at temperatures well below the melting temperature of NC cores. This idea has been used to facilitate transport measurements of arrays of cobalt NCs [10].

In this paper, we describe the electrical transport measured over the temperature range from 6 K to 250 K in three-dimensional close-packed arrays of organically capped CdSe NCs [2], as a function of annealing time and temperature. After the solids are annealed, three changes are observed. First, transmission electron micrographs (TEMs) confirm that the separation between NCs decreases with annealing. Second, the optical absorption spectrum changes: the excitonic peaks of the NC solids shift to lower energies and broaden with annealing. Both the TEM and the absorption data provide evidence that dots remain distinct after annealing. Last, the electronic properties of the solid change, and annealing results in greatly enhanced dark current and photocurrent. The dark current generally increases exponentially with voltage; we show fits to the form $I \approx V/R_o \, e^{|V|/V_o}$, where V is the voltage across the array of dots, and $R_o$ and $V_o$ are constants that depend on the properties of the array. We show that $R_o$ and $V_o$ both decrease as the solid is annealed, which corresponds to the measured current increasing by up to three orders of magnitude. Similar changes are observed in measurements of the photocurrent in annealed NC solids. This enhancement of the current cannot be explained by the observed decrease of inter-dot separation alone. Rather, it is likely that chemical changes in the inter-dot organic material during annealing reduce the barrier height. Our results suggest that controlled annealing can be a useful tool for improving the conduction properties of NC solids.

## II. EXPERIMENTAL DETAILS

We have prepared nanocrystals with diameters D from D = 4.3 nm to D = 6.1 nm, with the rms deviation $\sigma \approx 5\%D$ measured from optical absorption [2]. The NCs are capped by an organic molecule, tri-octylphosphine oxide (TOPO) which creates a large potential barrier for electrons moving between the NCs ($E_b \sim 1eV$ [8]), and whose length defines the surface-to-surface separation d = (1.1 ± 0.1) nm between nearest neighbor NCs [2]. Details of the nanocrystal synthesis may be found in Reference [2].

Figure 1 is a schematic of the typical device geometry used to study electrical properties of close-packed arrays of NCs, showing a film deposited on top of gold electrodes. The devices consist of microfabricated Au/Ti electrodes, 0.11 μm thick and 800 μm long, on insulating quartz substrates 0.5 mm thick. We have used separations between electrodes of L = 1 μm and L = 2 μm. Details of the device preparation are described in References [7,8]. The NC solids are formed by depositing a colloidal NC solution on the electrodes and the substrate and then drying it over several hours at room temperature in a nitrogen atmosphere. Most of the electrical measurements presented here correspond to a NC solid composed of approximately $10^9$ TOPO-capped CdSe NCs between gold electrodes L = 2 μm apart.  The thickness of the solid is ~ 0.4 μm (~ 70 monolayers), with a length corresponding to ~300 NCs between the electrodes.

For transport measurements, the CdSe NC-solids are annealed in vacuum inside the cryostat, up to temperatures $T_a$ = 110 C.  For $T_a$ > 110 C, samples are annealed in a separate system consisting of a hot plate in a forming gas atmosphere (95% nitrogen and 5% hydrogen).  They are then transferred back into the cryostat for electrical measurements.  Unless otherwise noted, the annealing time is 15 min. in forming gas and ≈ 1 hr in vacuum (~$10^{-6}$ torr).  We have annealed one transport sample in forming gas at these lower temperatures to ensure that its electrical transport properties are the same as for those annealed in vacuum.  For transmission electron microscopy (TEM) measurements, samples are deposited on carbon grids and for absorption measurements, samples are deposited on glass slides.  Both are annealed in forming gas.

Electrical transport measurements are performed in vacuum (p ≈ $10^{-6}$ torr) at temperatures from T = 6 K to T = 250 K and at applied electric fields up to E = 2.5x$10^8$V/m.  As illustrated in Fig. 1, a voltage V is applied across the electrodes and current I is measured with a current amplifier.  Control measurements on substrates without NCs ensure that substrate conduction is negligible up to T ~ 250 K.  The current detection limit in our measurements is ~ 5fA.  A green light-emitting diode (LED) is mounted ~3 mm above the device in the cryostat.  The peak wavelength of the LED is λ ≈ 565 nm, and the power P ≈ 50 μW.  This allows us to photo-excite NCs and to measure photocurrent as a function of V.

## III. RESULTS

The van der Waals attraction between NCs causes them to self-assemble into hexagonal close-packed arrays [2]. Transmission electron microscope (TEM) images provide direct evidence that annealing decreases the inter-dot separation in the NC solid. Figure 2 shows the transmission electron micrographs (TEMs) of arrays of CdSe NCs with D ≈ 6.1 nm as deposited (Fig.2 (a)), and after annealing at 350 C (Fig.2 (b)) and at 430 C (Fig.2 (c)). To estimate the average NC separation in Figs. 2(a) and 2(b) we have counted the number of NCs in the fixed sample area A = 63.5x63.5 nm$^2$ shown in Fig. 2. There are N ≈ 78 dots counted in this area before annealing, whereas there are N ≈ 94 dots after annealing at 350C. From the TEM data, the NC sizes after annealing at 350 C appear unchanged. Under this assumption and since N ∝ A/(D+d)$^2$ and d ≈ 1.1 nm [2] for the unannealed sample, we estimate that the inter-dot separation is reduced to d ~ 0.5 nm after annealing at 350 C. After heating at 430 C, the NCs are no longer distinct (Fig. 2 (c)).

In addition to sample characterization with TEM, optical absorption measurements have been used to characterize the distribution of NC sizes and to probe the coupling of NCs to their environment [2, 11]. Fig. 3 (a) shows the evolution of the absorption spectrum for one array of D ≈ 4.9 nm TOPO-capped CdSe NCs as deposited and after annealing at $T_a$ = 110 C, $T_a$ = 300 C, and $T_a$ = 350 C. In all the spectra, the main peak corresponding to the size-dependent band gap is clearly visible. As the annealing temperature is increased, the absorption spectrum shifts to the red and the peaks broaden. The presence of clear size-dependent excitonic features in the spectra of the annealed solids is additional evidence that the dots are still distinct. After annealing we observe an increased background in the absorption spectra, which has a long tail at long wavelengths. We infer from its wavelength dependence that it arises from scattering in the film. There may also be contributions from absorption due to pyrolyzing the organic material.

The band-edge peak at $\lambda \approx 587$ nm (E = 2.11 eV) in the as-prepared film, shifts by $\Delta E \approx -5$ meV, $-30$ meV, and $-40$ meV, after annealing at $T_a = 110$ C, $T_a = 300$ C, and $T_a = 350$ C, respectively. For reference, the band-gap of bulk CdSe corresponds to E = 1.8 eV ($\lambda \approx 690$ nm) [12], whereas the peak after the 350 C anneal is at 2.07 eV.

Fig. 3 (b) shows the evolution of the absorption spectrum for one array of smaller, $D \approx 4.3$ nm TOPO-capped CdSe dots as deposited and after annealing at $T_a = 350$ C for annealing times of t = 5, 10, 15 and 35 minutes. A subsequent annealing at $T_a = 400$ C for t = 10 min. is also shown. Curves 2 to 5 in Fig. 3 (b), indicate that the effect of annealing for longer times at the same temperature ($T_a = 350$ C) is to increase the width of the peak while affecting its position only slightly. The band-edge peak, which is at $\lambda \approx 566$ nm (E $\approx$ 2.19 eV) before annealing, has shifted by $\Delta E \approx -30$ meV after annealing at $T_a = 350$ C for 5 min. After subsequent annealing at $T_a = 350$ C for longer times, the band-edge peak shifted only by $\Delta E \approx -5$ meV from curve 2 to curve 5. After annealing at $T_a = 400$ C, the peak has shifted by $\Delta E \approx -290$ meV to E = 1.9 eV, making it very close to the gap of bulk CdSe.

From the TEM and absorption data, we have confidence that the dots are still physically distinct after annealing, as long as $T_a \leq 350$ C. We now describe how the transport properties of the NC solids change upon annealing. Figure 4 shows typical current-voltage (I-V) curves of a NC solid on log-linear plots at T = 77 K.  Measurements are shown both in the dark (Fig. 4 (a)) and with illumination from the LED (Fig. 4 (b)), for the film as-deposited (labeled I), and after annealing at $T_a$ = 110 C (II) and $T_a$ = 300 C (III).

We first focus on the dark-current data of Fig. 4 (a).  These clearly show the following features: (i) a hysteresis in the I-V curves; (ii) an approximately exponential increase of current with voltage, and (iii) an increased current through the solid after annealing.  We next discuss these features in turn.

(i) The hysteresis is such that the measured current is larger as the voltage sweeps up, from |V| = 0 to 500 V.  This hysteresis in the I-V curves is consistent with transient measurements, in which the current is observed to decrease as a power-law in time $I(t) = I_o t^{-\alpha}$, in response to a voltage step [7].  For the same samples studied here we have measured 30 minute long current transients I(t) in response to voltage steps, as explained in Reference [7], and from power-law fits we obtain $\alpha$ = (0.11± 0.04) independent of V.  One such current transient is shown in the inset of Fig. 4 (a).  Although there is no steady-state current, the small values of $\alpha$ correspond to extremely slow decays of current.  We find no significant change of $\alpha$ with annealing.  As shown in Fig. 4 (a) the current at 1s measured from the transient is in good agreement with that measured by sweeping V.

(ii)  The measured current is well described by an exponential form: we choose to fit to $I \approx V/R_o\, e^{|V|/V_o}$, in order to extract a characteristic voltage $V_o$ [V], and an equivalent resistance $R_o$ [$\Omega$] of the array.  From these exponential fits to the data in Fig. 4 (a), we obtain $V_o$ and $R_o$ for voltage sweeps up, from |V| = 0 to 500 V.  Before annealing

(curve I), we find $V_o \approx 31$ V and $R_o \approx 7 \times 10^{20}$ Ω. $R_o$ corresponds to a huge sheet resistance of $\sim 10^{23}$ Ω per square.

(iii) The measured current increases as the NC solid is annealed (curves II and III). The current after annealing at 300 C (III) is approximately three orders of magnitude larger than the original current (curve I) at fixed $|V| \approx 500$ V. The I-V curves after annealing are also steeper, corresponding to a decrease in $V_o$. From exponential fits to curves I to III, we find $V_o$ decreases from $\approx 31$ V to 23 V, and $R_o$ decreases from $\approx 7 \times 10^{20}$ Ω to $1 \times 10^{20}$ Ω.

The photocurrent through NC solids is also enhanced by annealing. Figure 4 (b) shows the corresponding current-voltage curves under constant excitation from the LED, before annealing (labeled I) and after annealing at temperatures $T_a = 110$ C (II) and $T_a = 300$ C (III). It appears that the photocurrent saturates at large electric fields ($E > 10^8$ V/m). This saturation of photocurrent has not been observed in previous measurements on as-deposited NC solids, presumably because lower fields have been used [8]. The origin of the saturation is being investigated in more detail [13]. At fields larger than $E \approx 2 \times 10^8$ V/m, the rise in the current appears to correspond to the onset of dark current (compare with Fig. 4 (a)). From Fig. 4 (b) we see that, like the dark current, the photocurrent increases after annealing. In the voltage range where the dark current is negligible ($|V| < 200$ V), the photocurrent increases by up to two orders of magnitude after the NC solid is annealed at $T_a = 300$ C. Because the LED is not monochromatic and because the intensity of light impinging on the sample is not well controlled, more detailed photocurrent measurements are being performed using laser excitation [13].

The I-V curves in Fig. 4 are presented as functions of voltage across the NC solid.

However, when I-V curves for NC solids with L = 1 μm and L = 2 μm are compared, we find that both the dark and photocurrent scale with the electric field $\varepsilon$ [7,8]. Specifically, the dark current follows I $\propto$ $e^{\varepsilon/\varepsilon_o}$, where $\varepsilon_o=V_o/L$, and $\varepsilon_o$ is independent of gap size L. We have confirmed this dependence for both the as-deposited and annealed samples.

The measurements described above have been performed on NC solids with different parameters, including gap size L (see Fig 1. (a)), NC diameter D, annealing temperature $T_a$ and time. The results are summarized for TOPO-capped NCs in Table I. Here $V_o$ is the exponent from the exponential fits to I-V curves for L = 2 μm. The other fit parameter $R_o$ is, to a first approximation, inversely proportional to the film thickness. It ranges from ~$10^{22}$ Ω to ~$10^{25}$ Ω per square for solids from ~1 μm to ~ 0.01 μm thick. The effects of annealing on transport do not depend on whether voltages are applied to the sample before annealing. We have also observed enhanced dark current and photocurrent for samples that have been stored at room temperature in vacuum (p ~ $10^{-6}$ torr) for several days. Finally, the results presented above are for NCs solids deposited on quartz substrates, but we have also used annealing to improve the conductivity of NC solids deposited on degenerately doped Si substrates with a thermally grown gate oxide ($SiO_2$).

### III. DISCUSSION

We have demonstrated the enhancement of the dark current and photocurrent in NC solids and the red-shifts and broadening of the absorption spectra. The dark current is found to be approximately an exponential function of the applied electric field, whereas the photocurrent has a more complicated functional form, and saturates at higher fields. In both cases, the functional forms of the I-V curves are approximately preserved after annealing, but the current increases by several orders of magnitude. In addition, TEM data show that the inter-dot separation decreases upon annealing.

The red-shift in the absorption spectra have been previously observed in NC systems as a function of the solvent composition [11], NC diameter and concentration [14,15]. These red-shifts have been attributed to the enhancement of either classical (dipole) coupling [11], or quantum-mechanical coupling between NCs [14,15]. To our knowledge, there are no previous observations of red-shifts caused by annealing.

As in Ref. [11], we find that the red-shifts observed in our work may be attributed to the change in polarization energy of the quantum-confined exciton when the dielectric environment around the NCs is changed. In the mean field approximation the average dielectric constant of the medium around each NC, $\varepsilon_m$, can be estimated as the volume-weighted average of dielectric constants for the organic spacer ($\varepsilon_r = 2.1$ [11]) and for the CdSe NC ($\varepsilon_r = 6.2$ [11]). When the inter-dot separation decreases, $\varepsilon_m$ increases because the volume fraction of CdSe increases. The average dielectric constant $\varepsilon_m$ can also increase because the dielectric constant of the organic material surrounding the NCs may increase with annealing. Correspondingly, the excitonic peak shifts to lower energies. This core-shell model of Ref. [11] can be applied in our case. From the TEM data we have found that the inter-dot separation decreases from $d = 1.1$ nm to $\sim 0.5$ nm after annealing at $\sim 350$ C, which here corresponds to the first excitonic peak shifting by $\Delta E \sim -40$ meV. Taking $\varepsilon_r = 2.1$ for TOPO and $\varepsilon_r = 6.2$ for CdSe NCs, the model of Eqs. [5] and [6] in Ref. [11] predicts that the change in NC spacing results in an energy shift of $\Delta E \sim -10$ meV, which is smaller than observed experimentally. Agreement is obtained only if one assumes that the dielectric constant of the organic cap increases slightly during annealing, from $\varepsilon_r = 2.1$ to $\varepsilon_r \sim 3$. Such a change in $\varepsilon_r$ might arise from chemical changes in the organic. Thus, the magnitude of the observed red-shifts could be understood to result from the change of the external dielectric environment.

We find only a weak temperature dependence of the dark current in the temperature range from T = 6 K to T = 250 K; in particular, the exponent $V_o$ in exponential fits of the I-V curves is approximately temperature independent. This weak temperature dependence suggests that electron tunneling is the main transport mechanism in CdSe NC solids for T < 250 K. References [6,7] provide evidence that the dark current is limited by electron tunneling from a space-charge region near the electron-injecting electrode into unoccupied dots. Similar behavior has been observed in linear chains of carbon particles for voltages lower than the threshold predicted for collective transport across the array [16]. To our knowledge, there is no theory to explain how the space-charge builds up in systems with long-range Coulomb interactions. However, as discussed below, the transport measurements shown here suggest that Coulomb correlations play a major role in the transport through CdSe NC solids.

The increased current after annealing is most likely caused by a combination of the decreased separation and the chemical transformation of the organic caps during annealing. When the width of the tunnel barrier is decreased by $|\Delta d|$, the tunneling probability is expected to increase by $\sim e^{\alpha |\Delta d|}$ ($\alpha^{-1} \sim 0.1$ nm for alkane molecules [8]). For annealing at 350 C, we have found $|\Delta d| \sim 0.6$ nm, so that the tunneling probability is expected to increase by a factor of $\sim 400$. This is consistent with results for the dark

current, but smaller than observed for the photocurrent. Previous photocurrent measurements [8] have been made with a different organic cap tributylphosphine oxide (TBPO), whose length is d = (0.7±0.1) nm [2], and whose barrier height is similar to that of TOPO. These measurements have shown a slightly smaller effect of decreasing inter-dot separation on transport properties than expected from a simple tunneling model when the barrier width is decreased. It is therefore likely that the changes in transport properties with annealing are caused both by the decreased separation and by the chemical transformation of the organic caps during annealing in a reducing atmosphere. TOPO decomposes into a carbonaceous substance during annealing and this process could lower the tunnel barriers between NCs.

Although the increase in dark current through the NC solids after annealing could result from the enhancement of inter-dot tunneling, a simple tunneling model cannot explain our dark current-voltage curves quantitatively. For the dark current measurements in Fig. 4 (a), we find an approximately temperature-independent exponential form $I \propto e^{|V|/V_o}$ of the current-voltage curves. Qualitatively, this form is expected for electron tunneling, and results from the lowering of inter-dot potential barriers when the electric field is applied across the NC solid. However, this simple model predicts a value for the coefficient $V_o$ which is inconsistent with the data. This can be shown by estimating the probability for electron tunneling $|T|^2$ from one NC to its nearest neighbor. In the semiclassical approximation,

$$|T|^2 \sim \exp[-2\int_0^d dx \sqrt{(2m^*/\hbar^2)(E_b - E_o - e\mathcal{E}x)}],$$ where $m^* \approx 0.1\ m_e$ [12] is the effective electron mass in CdSe, $\mathcal{E} \approx V/(\varepsilon_m L)$ is the electric field inside the NC solid, d is the

width of the barrier, $E_b$ is the height of the barrier, and $E_o$ is the ground state energy inside a NC. For small applied voltages per NC compared to the barrier height ($e\varepsilon d \ll E_b-E_o$), this expression reduces to $|T|^2 \sim \exp(\beta V)$, consistent with the form observed experimentally. However, $\beta = \sqrt{(2m^*/\hbar^2)(E_b - E_o)^{-1}}\ ed^2/(2\varepsilon_m L)$, and for d = 1.1 nm, L = 2 μm, $\varepsilon_m$ = 3.7 (for D = 5nm) [17], and assuming $E_b$-$E_o$ ~ 1 eV [8], we find that $V_o$ ~ 7 x $10^3$ V, whereas the measured value is ~ 30 V.

On the other hand, the characteristic voltage $V_o$ obtained from the exponential fits is comparable to the Coulomb interaction energies in the NC array. In the measurements on different sample lengths, we have found that the current depends on the electric field across the device. Therefore, for N dots across the gap, N ≈ L/(D+d), the exponential form of the dark current can be rewritten as $I \propto e^{v/v_o}$, where $v$ is the applied voltage difference across one NC in the array and $v_o = V_o/N$. We find that $v_o$ is comparable to the Coulomb interaction energy $E_{int}$ between charges on nearest-neighbor NCs. Specifically, from the data in Table I, we find $v_o$ ~ 50 to 300 meV. On the other hand, for charges on nearest-neighbor NCs embedded in a medium of average dielectric constant $\varepsilon_m$, the Coulomb interaction is $E_{int} = e^2/(4\pi\varepsilon_o\varepsilon_m(D+d))$. For D = 5 nm, d = 1.1 nm, and $\varepsilon_m \approx 3.7$ [17], we find $E_{int}$ ~ 60 meV, similar to $v_o$.

The changes of $v_o$ (=$V_o$/N) after annealing are comparable to the red-shifts ΔE of the band-edge peak. For example, in Fig. 3 (a) for D = 4.9 nm, we find ΔE ~ -30 meV after annealing at 300 C, whereas the corresponding change in $v_o$, found from the data of Fig. 4 (a), is Δ $v_o$ ~ - 20 mV. It seems reasonable to expect that these two quantities should be comparable if both of them reflect the Coulomb interaction between electron charges at a distance comparable to the NC diameter D. Specifically, we have proposed that the red-shifts reflect the Coulomb interaction energy of the electron-hole pair, whereas $v_o$ reflects the Coulomb interaction energy between electrons on nearest-neighbor NCs.

While conductance through arrays of metallic NCs has been studied theoretically, we know of no theoretical treatment of tunneling limited by long-range Coulomb interactions that might explain the exponential voltage dependence. In the absence of theoretical models, we find that the close agreement between $v_o$ and the Coulomb interaction energy $E_{int}$ between electrons on nearest-neighbor NCs strongly suggests the importance of Coulomb interactions in the tunneling current.

In conclusion, we have demonstrated that annealing of NC solids reduces the distance between NCs, and alters their optical and transport properties. Controlled annealing might serve as a useful tool to tune the properties of NC arrays and facilitate the observation of new collective phenomena.

This work has been supported by the MRSEC program of the National Science Foundation under award No. DMR 9808941. M.D. acknowledges support from the Pappalardo Fellowship.

**References:**

[1]  M.A. Kastner, Phys. Today 46, 24 (1993).

[2]  C.B. Murray, C.R. Kagan, M.G. Bawendi, Annu. Rev. Mater. Sci. 30, 545 (2000).

[3]  C. P. Collier, R.J. Saykally, J.J. Shiang, S.E Henrichs, J.R. Heath, Science, 277, 1978 (1997).

[4]  A.A. Middleton, N.S. Wingreen, Phys. Rev. Lett., 71, 3198 (1993).

[5 ]  L.S. Levitov, B. Kozinsky, cond-mat/9912484; D. S. Novikov, B. Kozinsky, L.S. Levitov cond-mat/0111345.

[6]  D.S. Ginger, N.C. Greenham, J. Appl. Phys. 87, 1361 (2000).

[7]  N.Y. Morgan, C.A. Leatherdale, M. Drndic, M. Vitasovic, M.A. Kastner, M.G. Bawendi, submitted (2001).

[8]  C.A. Leatherdale, C.R. Kagan, N. Y. Morgan, S.A. Empedocles, M.A. Kastner, M.G. Bawendi, Phys. Rev. B 6303, 9901 (2001).

[9]  W. Xue, P.A Lee, Phys. Rev. B 38, 9093 (1988).

[10]  C.T. Black, C. B. Murray, R.L. Sandstrom, S.H. Sun, Science 290, 1131 (2000).

[11]  C.A. Leatherdale, M.G. Bawendi, Phys. Rev. B 6316, 5315 (2001); See also, M. Iwamatsu, M. Fujiwara, N. Happo, and K. Horri, J. Phys.: Condens. Matter 9, 9881 (1997).

[12]  Landolt H., Bornestein R., *Numerical data and functional relationships in science and technology*, Springer-Verlag, Berlin, 1961.

[13]  M. Vitasovic et al., to be submitted.

[14]  M.V. Artemyev, A. I. Bibik, L. I. Gurinovich, S. V. Gaponenko, U. Woggon, Phys. Rev. B 60, 1504 (1999).

[15] O.I. Micic, S.P. Ahrenkiel, A.J. Nozik, Appl. Phys. Lett. 78, 4022 (2001).

[16] A. Bezryadin, R.M. Westervelt, M. Tinkham, Appl. Phys. Lett. 74, 2699 (1999).

[17] Assuming a close-packed 3D array of TOPO-capped CdSe dots: $\varepsilon_m = 2 + 16.8\sqrt{2}\pi/3(R/(2R+d))^3$, where R is the dot radius and d is the surface-to-surface separation.

**Figure Captions:**

Fig.1. Schematic of the device used to study the electrical properties of close-packed CdSe NC solids. Gold electrodes, separated by distances of L = 1 μm and L = 2 μm, are microfabricated on quartz substrates. NC diameters vary from D = 4.3 nm to D = 6.1 nm in this work.

Fig.2. Transmission electron micrographs of CdSe NC solids with diameter D = 6.1 nm: (a) as deposited, and after annealing in forming gas at (b) $T_a$ = 350 C and (c) $T_a$ = 430 C. The inter-dot separations found from these data are d ≈ 1.1 nm for (a) and d ≈ 0.5 nm for (b). The NCs are no longer distinct at 430 C.

Fig.3. Absorption spectra of CdSe NC solids deposited on glass slides before annealing (curve 1) and after annealing in forming gas as a function of (a) annealing temperature $T_a$, for D = 4.9 nm and (b) annealing time $t_a$, for D = 4.3 nm. The absorption spectra are shifted vertically for clarity and the positions of the band-edge peaks are indicated. (a) The curves labeled 1-4 are as-deposited, and then after annealing at $T_a$ = 110 C, 300 C, and 350 C, respectively. (b) The curves labeled 1-5 are as prepared and annealed at 350 C for $t_a$ = 5, 10, 15, 35 min., respectively. The dashed curve is for the same film after annealing at $T_a$ = 400 C for 10 min.

Fig.4. Current-voltage curves of a CdSe NC solid at T = 77 K (a) in the dark, and (b) during illumination with a green LED. Curves labeled I-III are for the as prepared film,

and after annealing at $T_a = 110$ C and $T_a = 300$ C, respectively. The NC diameter is D = 4.9 nm for the film used. In (a) the solid lines represent voltage sweeps at a rate $\Delta V/\Delta t \approx 1$ V/s from V = 0 to positive voltages and back to V = 0, and then symmetrically to negative voltages and back to V = 0. Triangles ($\Delta$) and circles ($\circ$) are the amplitudes of the current at 1s after applying a voltage step (V = 0 for t < 0, V = $V_{el}$ for t $\geq$ 0). The dashed line represents the exponential fit for curve III. Inset: One current transient, which is well-described by $I(t) \approx 78\, t^{-0.13}$ pA (t is in seconds), for an applied voltage step at V = - 460 V after annealing at 110 C.

Tab. I: Summary of results for TOPO-capped CdSe NC solids as a function of NC diameter, D, and annealing temperature $T_a$: $V_o$ is the exponent from fits to I-V curves for L = 2 µm, as described in the text.

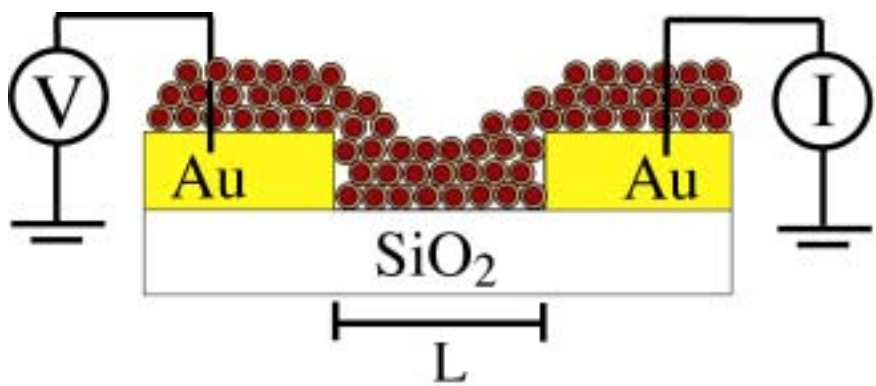

Figure 1

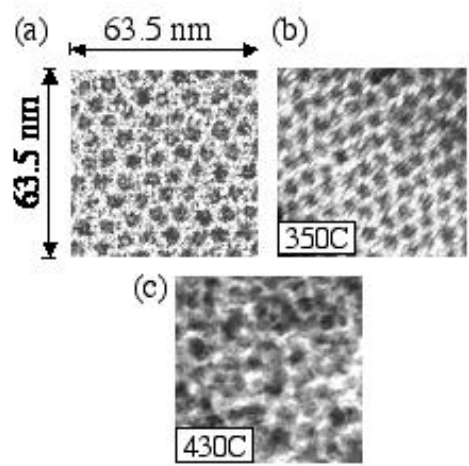

Figure 2

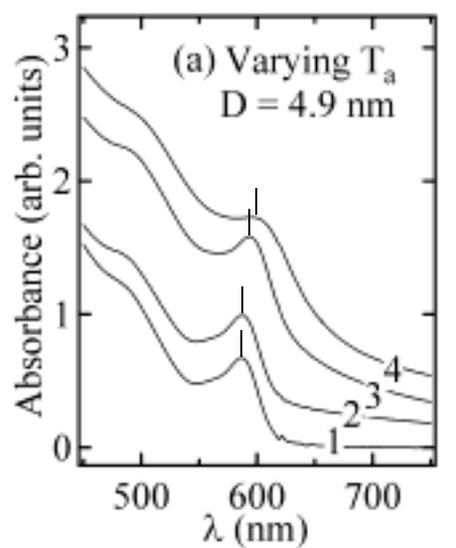
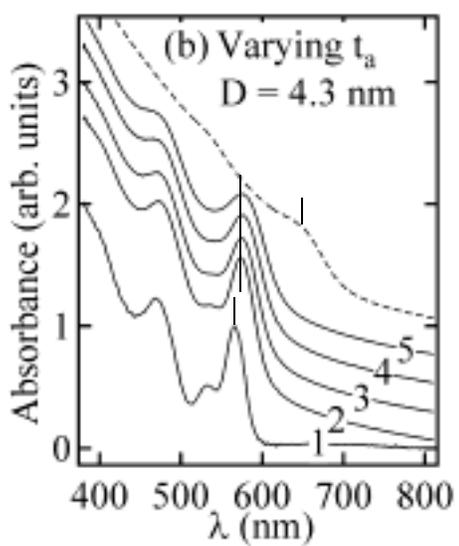

Figure 3

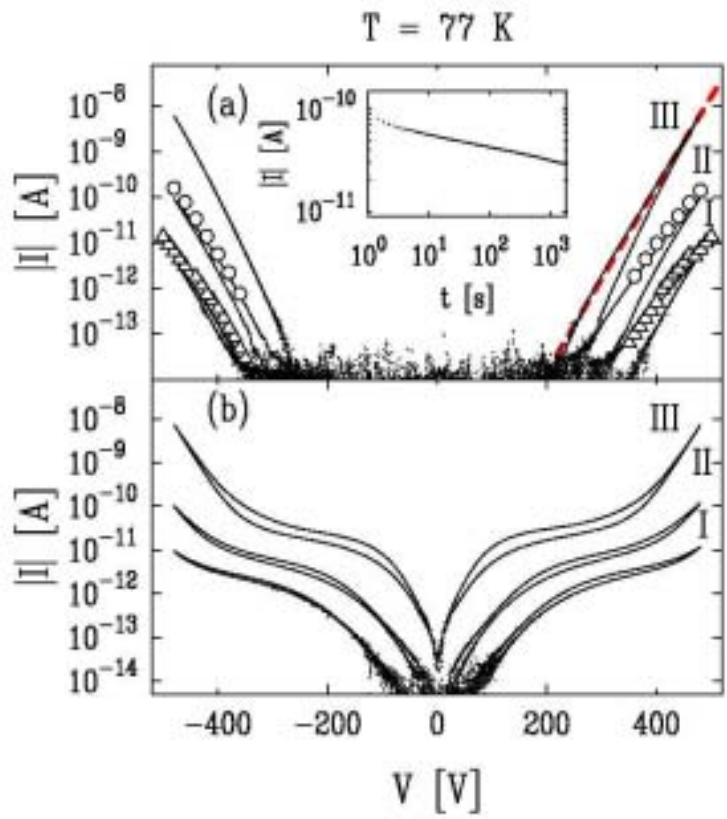

Figure 4

TAB. I

| Sample | D (nm) | $T_a$ (C) | $V_o$ (V) |
|--------|--------|-----------|-----------|
| A | 6.1 | RT | 78±1 |
| A | 6.1 | 350 | 32±5 |
| B | 4.9 | RT | 31±4 |
| B | 4.9 | 110 | 28±2 |
| B | 4.9 | 300 | 23±5 |
| C | 4.3 | 350 | 21±1 |